\documentclass[a4paper]{jpconf}
\usepackage{graphicx}
\usepackage{units}
\usepackage[pdfpagelabels,bookmarks=true,colorlinks,linkcolor=black,urlcolor=black,citecolor=blue]{hyperref}
\newcommand{\Mac}{MAC-E filter}
\begin{document}
\title{The KATRIN Experiment}

\author{Marcus Beck for the KATRIN collaboration}

\address{Institut f\"ur Kernphysik, Westf\"alische Wilhelms-Universit\"at M\"unster, Germany}

\ead{marcusb@uni-muenster.de}

\begin{abstract}
The KArlsruhe TRitium Neutrino mass experiment, KATRIN, aims to search for the mass of the electron neutrino with a sensitivity of $\unit[0.2]{eV/c^2}$ (90\% C.L.) and a detection limit of $\unit[0.35]{eV/c^2}$ ($5 \sigma$). Both a positive or a negative result will have far reaching implications for cosmology and the standard model of particle physics and will give new input for astroparticle physics and cosmology. The major components of KATRIN are being set up at the Karlsruhe Institut of Technology in Karlsruhe, Germany, and test measurements of the individual components have started. Data taking with tritium is scheduled to start in 2012.
\end{abstract}

\section{Introduction}

After neutrino oscillation experiments have shown that neutrinos possess a non-zero rest mass the question of the absolute mass scale of neutrinos has become important for particle physics and cosmology (see e.g. \cite{otten-weinh-review,lesgourgues-pastor-review}). The goal of the KArlsruhe TRItium Neutrino experiment (KATRIN, \cite{kdr}) is to search for the mass of the electron antineutrino with a sensitivity of $\unit[0.2]{eV/c^2}$. This will probe most of the mass range in which the three neutrino flavours have nearly degenerate masses and where neutrinos are of cosmological importance.

To this end a precision measurement of the endpoint region of the $\beta$-decay of tritium will be performed with KATRIN. The shape of this spectrum depends sensitively on the neutrino mass. This will be done by using a windowless gaseous tritium source for small systematic uncertainties, and an electrostatic energy filter of {\Mac} type (electrostatic filter with magnetic adiabatic collimation, \cite{picard-nimb,lobashev85}) for the analysis of the electron energy with high luminosity and high resolution. This is the same method as used at the Mainz \cite{picard-nimb,Kra05} and Troitzk neutrino mass experiments \cite{lobashev85,Lob99}, which have set the until now best upper limits of  $\unit[2.3]{eV/c^2} (95\%)$ on the neutrino mass.

\section{The set-up of KATRIN}

\begin{figure}
\begin{center}
\includegraphics[width=\textwidth]{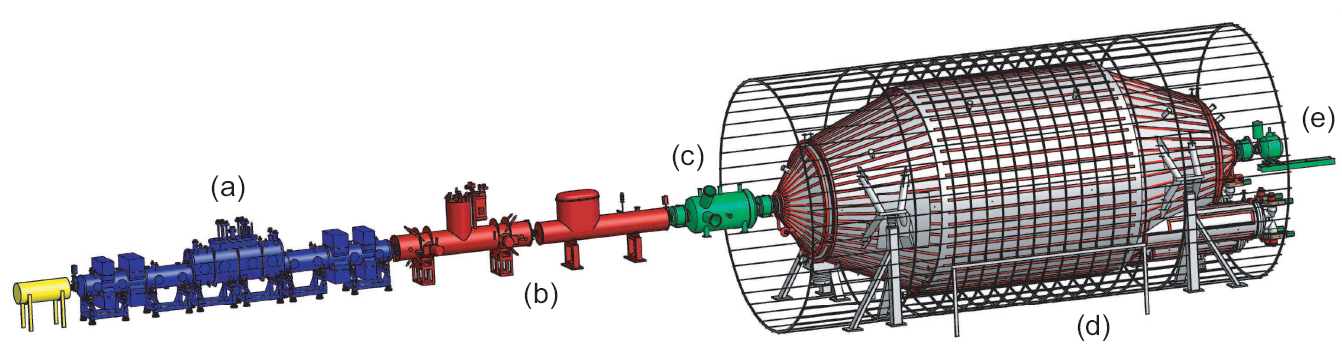}
\end{center}
\caption{\label{fig:setup}The KATRIN set-up (explanation see text)}
\end{figure}

Figure~\ref{fig:setup} gives an overview of the KATRIN set-up. It consists of a source section with the windowless gaseous tritium source, the WGTS (a), a transport section with a differential- and a cryopumping section (b), a pre-spectrometer (c), the main spectrometer (d) and a detection section with the focal plane detector (e).

The WGTS provides a source of electrons from tritium $\beta$-decay with high and uniform intensity. It basically consists of a $\unit[10]{m}$ long beam tube of $\unit[90]{mm}$ diameter through which tritium is circulated in a loop with a flow of $\unit[4.7]{Ci/s}$ and a maximum pressure of $\unit[4 \cdot 10^{-3}]{mbar}$. This results in a column density of $\unit[5 \cdot 10^{17}]{T_2/cm^2}$ and $\approx 10^{10}$ $\beta$-particles per second accepted by the spectrometer. In order to keep density fluctuations at $0.1\%$ or less the WGTS is temperature stabilized to $\unit[30]{mK}$ at a temperature of $\unit[30]{K}$ by two-phase neon cooling. The purity of the $T_2$ is maintained at $95 \pm 1 \%$ by continous purification of the gas in the tritium loop. These stringent stability requirements result in a highly complex cryostat system \cite{Gro09,Bor06}. The technologies used will be tested using a demonstrator set-up, which will be available in spring 2010. In the final set-up the properties of the source will be checked regularly with an electron source, which is under development \cite{Val09a}, by laser Raman spectroscopy, which already has been deployed successfully for test measurements \cite{Stu10}, and a Penning trap in the beamline to the spectrometers using FT-ICR, which has recently been tested successfully \cite{Ubi09}.

The transport section serves to guide the electrons from tritium $\beta$-decay to the spectrometers without any losses and to prevent the transport of tritium at the same time, with a reduction factor for tritium of $10^{14}$ and a final flow of $10^{-14}$mbar-l/s. This is achieved by a differential pumping section, which recently arrived at Karlsruhe for testing and commissioning, followed by a cryogenic pumping section. Here the remaining tritium gets cryosorbed on a layer of Argon frost on the walls of the stainless steel vessel at a temperature of $\approx \unit[3.4]{K}$. This method has been tested with the TRAP experiment successfully \cite{Kaz08,Eic08}.

The pre-spectrometer is a \Mac \ like the main spectrometer. It rejects all electrons with energy $< \unit[200]{eV}$ below the endpoint, which are of no interest for the neutrino mass measurement and which even may increase the main spectrometer background. It has been successfully tested and is working at design parameters with an intrinsic background of $\cal{O}$($10^{-2}$ counts per second (cps)). For now it is being used as a test case for the technologies and methods used for the main spectrometer, such as the vacuum system (see e.g. \cite{Luo07,Wol09}) and the electromagnetic design. For these tests it is presently equipped with a 64-pixel PIN diode \cite{Wue06}.

The endpoint region of the $\beta$-spectrum is measured with the main spectrometer. This consists of a stainless steel vacuum vessel of $\unit[23]{m}$ length and $\unit[10]{m}$ diameter and will reach a vacuum of $\unit[10^{-11}]{mbar}$ using turbomolecular pumps and non-evaporating getter material. The magnetic field at the entrance of this \Mac \ is $\unit[4.5]{T}$ and at its exit $\unit[6]{T}$ while the field in the central analysis plane will be $\unit[3 \cdot 10^{-4}]{T}$ resulting in an energy resolution of $\unit[0.93]{eV}$. The retardation voltage is applied directly to the spectrometer hull. For fine shaping of the field and background suppression a modular inner wire electrode system will be used.
The main spectrometer has been set-up at Karlsruhe and an outgassing rate of $<10^{-12}$mbar-l/s has been achieved after bakeout. High demands on the quality of the electric potential and therefore for the precision of the machining of the wire modules required stringent quality control (see e.g. \cite{Pra09}). The installation of the wire modules is in progress.

The high voltage (HV) that creates the retardation potential in the analysis plane has to be highly stable. To reach the sensitivity of KATRIN, fluctuations of the HV of $\unit[3]{ppm}$, i.e. of $\unit[60]{mV}$, have to be detected. For this purpose a precision HV-divider has been developed \cite{Thu09} to measure the applied voltage with this precision. In conjunction with calibration sources based on conversion electrons from the decay of $^{83m}Kr$ this will ensure the required short and long term stability of the high voltage.

Electrons with sufficient energy to pass the retardation potential are counted with the focal plane detector. This is a 148-fold segmented PIN diode with surroundings made of low-activity materials to reduce the intrinsic background of the detector. Simulations show that a background of $\unit[3 \cdot 10^{-3}]{cps}$ can be achieved in the signal region. The detector will be commissioned at Karlsruhe in spring 2010.

The KATRIN background count rate necessary to reach the intended sensitivity is $\unit[1 \cdot 10^{-2}]{cps}$. By shifting the signal in the detector via a DC-voltage the intrinsic detector background can be reduced to $\unit[1 \cdot 10^{-3}]{cps}$. However, other sources of background have to be controlled as well. Two major sources are background from the tank walls due to muon interactions and radioactive decays, and background from incidential Penning traps due to the magnetic and electric fields employed in the \Mac. The former is suppressed by the axial magnetic field, which guides most electrons from the wall back to the wall. The remaining wall electrons will be suppressed by the wire electrode system, which is slightly more negative (O(100V)) than the tank wall. The background due to Penning traps has been investigated in detail at the pre-spectrometer and can be avoided in large part by proper shaping of the electrodes. However, some unavoidable Penning traps still exist, like the trap formed by the pre- and the main spectrometer. Investigation of these traps has started and first results are promising \cite{Bec09}.

\section{Conclusion}

The KATRIN experiment searching for the neutrino mass is under construction. The major components are being assembled and tested, and integration of the components will start soon. Installation of the inner electrode system of the main spectrometer is ongoing. In 2010 the properties of the main spectrometer will be investigated using calibration sources. Data taking using tritium will start in 2012 and will proceed for five years.

\section{Acknowledgement}

The work by the author is supported by the German Bundesministerium f\"ur Bildung und Forschung.

\end{document}